\documentclass[namedreferences]{SolarPhysics}
\usepackage[optionalrh,solaenum]{spr-sola-addons} 
\usepackage{graphicx}                    
\usepackage{color}                       
\usepackage{url}                         

\usepackage{subfig}
\usepackage{multirow}
\usepackage{caption}

\newcommand{\aap}{Astron. Astrophys.}
\newcommand{\ag}{Ann. Geophys.}
\newcommand{\apj}{Astrophys. J.}
\newcommand{\solphys}{Solar Phys.}
\newcommand{\jgr}{J. Geophys. Res}
\newcommand{\ssr}{Space Sci. Rev.}
\newcommand{\grl}{Geophys. Res. Lett.}
\newcommand{\apjl}{Astrophys. J. Lett.}
\newcommand{\adv}{Adv. Space Res.}

\newcommand{\Rsolar}{\mbox{$\rm R_{\odot}\:$}}
\newcommand{\Rsol}{\mbox{$\rm R_{\odot}$}}
\newcommand{\VR}{$V\hskip-0.0375cmR\:$}
\newcommand{\VRR}{$V\hskip-0.0375cmR$}
\newcommand{\QVR}{$Q\mbox{(\VRR)}\,$}
\begin{document}

\begin{opening}

\title{Variation of proton flux profiles with the observer's latitude in simulated gradual SEP events}

%
%
\author{R.~\surname{Rodr\'{\i}guez-Gas\'{e}n}$^{1,2,3}$\sep
        A.~\surname{Aran}$^{3,4}$\sep
        B.~\surname{Sanahuja}$^{3}$ \sep C.~Jacobs$^{5,6}$ \sep S.~Poedts$^{5}$}


\runningauthor{R.~Rodr\'{\i}guez-Gas\'{e}n \textit{et al.}}
\runningtitle{Variation of proton flux profiles with the observer's latitude in simulated gradual SEP events}


    \institute{$^{1}$ LESIA, Observatoire de Paris, CNRS, UPMC and Univ.~Paris-Diderot, F-92195 Meudon Cedex, France.\\ Email: \url{rosa.rodriguez@obspm.fr}\\
    	           $^{2}$ CSNSM, IN2P3-CNRS, Univ.~Paris-Sud, F-91405 Orsay Cedex, France\\             
    		  $^{3}$ Departament d'Astronomia i Meteorologia and Institut de Ci\`{e}ncies del Cosmos, Universitat de Barcelona, E-08028 Barcelona, Spain\\ 
                     $^{4}$ Research and Scientific Support Department of ESA, ESTEC, 2200 AG Noordwijk, The Netherlands\\
		  $^{5}$ Centrum voor Plasma-Astrofysica and Leuven Mathematical Modelling Centre, K.U.Leuven, 3001 Leuven, Belgium\\
		  $^{6}$ Now working at Space Applications Services N.V., B-1932 Zaventem, Belgium}
\begin{abstract}
We study the variation of the shape of the proton intensity-time profiles in simulated gradual Solar Energetic Particle (SEP) events with the relative observer's position in space with respect to the main direction of propagation of an interplanetary (IP) shock. Using a three-dimensional (3D) magnetohydrodynamic (MHD) code to simulate such a shock, we determine the evolution of the downstream-to-upstream ratios of the plasma variables at its front. Under the assumption of an existing relation between the normalized ratio in speed across the shock front and the injection rate of shock-accelerated particles, we model the transport of the particles and we obtain the proton flux profiles to be measured by a grid of 18 virtual observers located at 0.4 and 1.0\,AU, with different latitudes and longitudes with respect to the shock nose. The differences among flux profiles are the result of the way each observer establishes a magnetic connection with the shock front, and we find that changes in the observer's latitude may result in intensity changes of up to one order of magnitude at both radial distances considered here. The peak intensity variation with the radial distance for the pair of observers located at the same angular position is also derived. This is the first time that the latitudinal dependence of the peak intensity with the observer's heliocentric radial distance has been quantified within the framework of gradual SEP event simulations.
\end{abstract}
\end{opening}
\section{Introduction}\label{s1}

Solar radiation storms caused by Solar Energetic Particle (SEP) events represent one of the most severe risks in space environment \cite{Feynman00,Cucinotta10,Zeitlin13}. Hence, in terms of the launch and operation of space vehicles, manned or unmanned exploration of the inner solar system, as well as in terms of the societal impact, it is crucial to be able to predict their occurrence and their intensity. Nevertheless, at present, model predictions are unable to reproduce observations of gradual SEP events with enough accuracy (see the report of the National Research Council in 2008\footnote{~\url{www.nap.edu/catalog.php?record_id=13060}} and \opencite{Luhmann10}). The main reasons are both the lack of \textit{in situ} information on the actual environment where SEP events develop (most observations are single or double point measurements), and the simplified assumptions made in the models to render the event simulations tractable  (see more comments on this issue in Section~1.2 of \opencite{RG11}).

The variation of SEP event characteristics with the observer's heliolatitude has been studied using data from \textit{Ulysses} (\opencite{Lario01}, 2005; \opencite{Simnett01}; \opencite{Dalla03a},b; \opencite{Dalla10}). Its influence, however, has not been quantitatively considered yet in simulations of gradual SEP events, mainly because the majority of magnetohydrodynamic (MHD) models used until now to simulate the coronal and interplanetary (IP) associated shocks involve only 1D, 2D or 2.5D codes (see, \textit{e.g.}, \opencite{Lario98}; \opencite{Aran07b}; \opencite{Verkhoglyadova10}, 2012). The scarce number of attempts to simulate SEP events using 3D shock propagation models have been applied to near-ecliptic SEP observations at 1.0\,AU (with the observer located not so far away from the central part of the IP shock; \textit{e.g.} \opencite{Roussev04}; \opencite{Sokolov04}; \opencite{Luhmann10}; \opencite{Rouillard11}). The number of studies involving the modelling of multi-spacecraft SEP events observed at different heliographic latitudes is also small (\textit{i.e.}, \opencite{Dayeh10}; \opencite{Agueda12}) and the simulation of a moving source of particles is not considered.

In the context of numerical simulations, not involving the modelling of a specific observed event, \inlinecite{RG10} showed that two observers located at the same radial distance, with the same heliolongitude, and magnetically connected to the same IP shock, would not necessarily observe the same particle flux profile if they are at different latitudes (-15$^{\circ}$ to +15$^{\circ}$) with respect to the incoming shock. The reason is that their magnetic connection with the shock front may scan different regions with different conditions for particle shock-acceleration and, hence, the observed flux profiles may largely differ. Consequently, 3D models able to simulate the shock propagation out of the ecliptic plane are essential to reliably reproduce the main features of a SEP event. 

Moreover, the history of the evolution of the shock variables near the Sun is required to reproduce the prompt phase of many gradual events (\textit{e.g.} \opencite{Aran08}; \opencite{RG10}). This means that the propagation of the IP shock must be simulated from the vicinity of the Sun. Then, shock simulations starting close to the Sun aim at addressing the evolution of the shock strength at the first stages of the SEP event, depending on the spacecraft's magnetic connection with respect to the solar parent activity location (\textit{e.g.} \opencite{Kocharov12}).

In this work we discuss the synthetic flux profiles obtained for several observers when applying the \QVR relation\footnote{~The \QVR relation is a semi-empirical relation between the injection rate of shock-accelerated particles, $Q$, and the normalized radial velocity jump across the shock front, \VR (see later); more details can be found in \inlinecite{Lario98}, \inlinecite{Aran07a}, and \inlinecite{Aran07b}.} used to build the Solar Particle Engineering Code (SOLPENCO) tool \cite{Aran06}. Using a simulation of an IP shock we have, first of all, analysed the main features of several shock variables at the point of the shock where the observer is magnetically connected to (\textit{i.e.}, the cobpoint, \opencite{Heras95}). To simulate the generation and propagation of the shock, we have used the 3D MHD model developed by \inlinecite{Jacobs07b}. We have followed its evolution up to a set of 18 virtual observers situated at 0.4\,AU and 1.0\,AU, and at different heliolongitudes and heliolatitudes, extending in this way our previous work \cite{RG10}. We have obtained the evolution of the plasma variables and magnetic field, as well as of the shock strength at the cobpoint (characterized by the normalized radial velocity jump, \VRR), for each observer. Then, using the \QVR relation of SOLPENCO, we have derived synthetic flux profiles and we have studied the radial variation of peak intensities for the set of observers. Our main goal is to study the variation of the proton flux profiles to be measured by each observer due to their different magnetic connections with the shock front.

We present the evolution of plasma variables and magnetic field, and the particle flux profiles, as if they were real events detected by several observers located at different radial distances from the Sun, and different longitudes and latitudes. It does not exist any observed gradual SEP event out of the ecliptic plane of similar characteristics as the ones simulated here; thus, the obtained results cannot be directly compared with observations. We stress the fact that for incoming space missions, as \textit{Solar Orbiter}, \textit{Beppi Colombo} and \textit{Solar Probe Plus}, these type of 3D models will be necessary to estimate and predict SEP events\footnote{~As stated by an expert committee under the auspices of the National Research Council during the latest decadal survey of the space weather field; \url{www.nap.edu/catalog.php?record_id=13060}.}.

The outline of the article is as follows. In Section~\ref{s2} we shortly summarize the main features of the shock propagation model, the IP scenario and how the cobpoint and the plasma jumps at the shock front are determined. In Section~\ref{s3} we present the results of the simulation concerning the evolution of the plasma and magnetic field variables, and the radial velocity jump. Section~\ref{s4} shows the derived synthetic flux profiles and how the peak intensities vary with the observer's radial distance from the Sun. Conclusions of this work are given in Section~\ref{s5}.

\section{The model}
\label{s2}

The model used to simulate the shock propagation and the location of the cobpoint is fully described in \inlinecite{RG10} and \inlinecite{RG11}. Here we review the principal characteristics.

\subsection{Background solar wind and shock simulation} 
\label{s2.1}

We simulate the 3D background solar wind using the model developed by \inlinecite{Jacobs07b}. The left panel of Figure~3.1 of \inlinecite{RG11} shows, from top to bottom, the magnetic field strength,  $B$, solar wind density, $n$, and radial velocity, $\upsilon_{r}$, profiles as a function of the radial distance from the Sun for three different latitudes (7$^{\circ}$, 22$^{\circ}$ and 45$^{\circ}$ above the solar equator). In this figure, a fast regime ($\sim$\,700\,km\,s$^{-1}$) at high latitudes and a slow regime ($\sim$\,400\,km\,s$^{-1}$) near the equator can be clearly differentiated. Table~\ref{table1} summarizes the derived values of the magnetic field strength, $B$, solar wind density, $n$, and radial velocity, $\upsilon_{r}$, at 0.4\,AU and at 1.0\,AU, and at latitudes 7$^{\circ}$, 22$^{\circ}$ and 37$^{\circ}$ above the solar equator, respectively. It can be noted that: (1) the solar wind speed values reproduce the typical observed values at different latitudes and radial distances \cite{Volkmer85,Balogh08}; (2) at 0.4\,AU, $B$ varies in latitude from 7\,nT to 11\,nT, underestimating observations near the ecliptic plane at such distance \cite{Mariani90}; and (3) the solar wind density is relatively high in comparison to observations. For example, at 1\,AU in the ecliptic plane, the model gives $\sim$\,35\,cm$^{-3}$ whereas the observed average value is $\sim$\,6\,cm$^{-3}$ \cite{Kivelson95}.

For the simulation of a fast shock, we use the same density-driven pulse to trigger a disturbance as in \inlinecite{RG10}. The shock is generated by superimposing a high-density moving plasma blob on the background solar wind. We assume that the plasma bubble has a radius $d_{CME}$\,=\,0.75\,\Rsolar and that it is centered at 2.5\,\Rsolar from the Sun. The extra density and velocity introduced in the plasma blob are: $n_{CME}$\,=\,7\,$\times$\,10$^{8}$\,cm$^{-3}$ and $\upsilon_{CME}$\,=\,3500\,km\,s$^{-1}$. These values yield a total mass of $\sim$\,1.4\,$\times$\,10$^{17}$\,g and a kinetic energy of $\sim$\,2.5\,$\times$\,10$^{33}$\,erg (both quantities within the range of estimated values for fast coronal mass ejections (CMEs); \opencite{Colaninno09}). The blob is launched in the direction ($\theta_{CME}$\,=\,22$^{\circ}$,\,$\varphi_{CME}$\,=\,142$^{\circ}$); see Figure~2.4 of \inlinecite{RG11}.

The shock simulation is performed using a co-rotational 3D MHD model in spherical coordinates ($r$,\,$\theta$,\,$\varphi$). The computational domain covers 1\,$\leq$\,$r$\,$\leq$\,220\,\Rsolar\\ in radial direction, and -90$^{\circ}$\,$\leq$\,$\theta$\,$\leq$\,90$^{\circ}$ and 0$^{\circ}$\,$\leq$\,$\varphi$\,$\leq$\,360$^{\circ}$ in angular directions. The radial step varies from $\delta r$\,=\,0.02\,\Rsolar near 1\,\Rsolar to $\delta r$\,=\,0.24\,\Rsolar at 30\,\Rsolar and further on. The grid in $\theta$-direction varies from a maximum angular step, $\delta \theta$\,=\,3.88$^{\circ}$, near the poles to a minimum value,  $\delta \theta$\,=\,0.8$^{\circ}$, at the equator. In $\varphi$-direction the grid is uniform, with $\delta \varphi$\,=\,2$^{\circ}$. The ideal MHD equations describing the evolution of the perturbation are solved by means of the Total Variation Diminishing Lax-Friedrichs scheme. The maintenance of a vanishing divergence of the magnetic field down to machine round-off error is guaranteed at all time by using the vector potential at the nodes according to the constrained transport method \cite{Evans88}. 

\begin{center}
\begin{table}[t!]
\caption{Values of $B$, $n$ and $\upsilon_{r}$ of the simulated solar wind at 0.4\,AU (left) and at 1.0\,AU (right), for the three latitudes.}
\label{table1}
\begin{tabular}{ccccccccc}
\hline
 & & \multicolumn{3}{c}{0.4\,AU} & & \multicolumn{3}{c}{1.0\,AU}\\
  & $\theta$\,[$^{\circ}$] & 7 &  22 & 37 & & 7 & 22 & 37\\
\hline
$B$\,[nT] & & 10.9 & 11.8 & 7.2 & & 2.4 & 2.3 & 1.2\\
\noalign{\smallskip}
$n$\,[cm$^{-3}$] & & 209.8 & 164.3 & 108.2 & & 34.3 & 25.3 & 17.0\\
\noalign{\smallskip}
$\upsilon_{r}$\,[km\,s$^{-1}$] & & 381 & 428 & 638 & & 383 & 440 & 651\\
\hline
\end{tabular}
\end{table} 
\end{center}

\subsection{Location of the observers in space} 
\label{s2.2}

The evolution of the simulated shock is followed from 4\,\Rsolar up to 220\,\Rsol, with nine observers located at $r$\,=\,86\,\Rsolar ($\sim$\,0.4\,AU, approximately Mercury's orbit), and other nine located at 215\,\Rsolar ($\sim$\,1.0\,AU). 

Since the simulated shock is travelling very fast, it is not geometrically deflected by the background medium. Thus, the leading edge of the shock, the so-called nose of the shock, is moving along the same direction as the launched direction. In longitude we have placed three observers 45$^{\circ}$ westward of the nose of the shock. That is, these observers would see the solar event as a W45 event. Three other observers are located 30$^{\circ}$ eastward of the shock nose, and  three more in the same longitude as the shock direction (central meridian observers). Hence, looking to the Sun, these observers are placed at W45, E30 and W00 with respect to the CME launched direction. In latitude, three observers are placed at the same latitude as the shock leading direction, $\theta$\,=\,22$^{\circ}$, which are the N22 observers. Three other observers are placed 15$^{\circ}$ northward of this direction, being the N37 observers ($\theta$\,=\,37$^{\circ}$), and three more are located 15$^{\circ}$ southward, these are the N07 observers ($\theta$\,=\,7$^{\circ}$). 

\begin{figure}[!ht]
\centering
\subfloat{\includegraphics[width=0.675\textwidth]{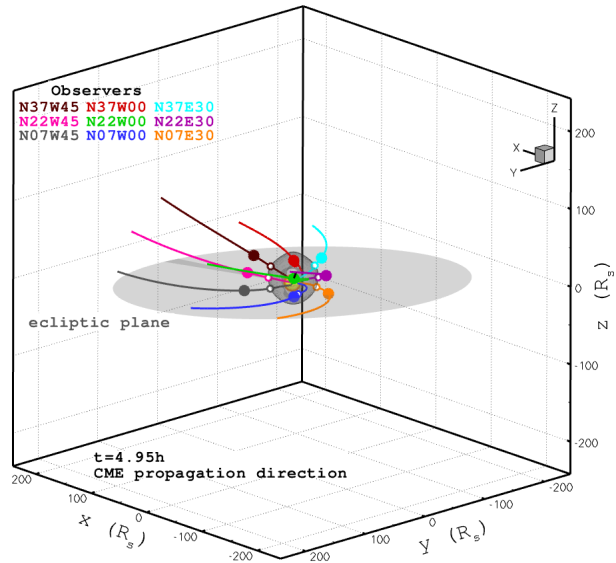}}
\caption{3D frontal view of a snapshot of the 3D simulation after t = 4.95\,hours (adapted from Figure~2 of \protect \opencite{RG10}). The dark-grey isosurface indicates the location of the shock front. The grey slice shows the ecliptic plane, the yellow circle marks the position of the Sun and the black line indicates the CME-driven shock propagation direction (towards the reader visual, nearly hidden by the green observer). The colour code for the nine observers, their corresponding IMF lines and cobpoints are as labelled.}
\label{figobs3d1}
\end{figure}

In short, the observers are located at the angular positions N37W45, N22W45, N07W45, N37W00, N22W00, N07W00, N37E30, N22E30 and N07E30, with the shock launched in the N22W00 direction. Figure~\ref{figobs3d1} shows a 3D frontal view of one snapshot of the shock simulation 4.95\,hours after the launch of the perturbation, as well as the location of the nine 0.4\,AU-observers; the angular positions for the 1.0\,AU-observers are the same. The dark-grey surface indicates the considered spatial boundary of the front of the expanding IP shock (see Section~\ref{s2.3}). The black line indicates the CME propagation direction. Coloured solid circles indicate the position of the observers, the colour lines their interplanetary magnetic field (IMF) lines and the coloured open circles the location of their corresponding cobpoints. The colour code for the nine observers is as labelled (see more details in \opencite{RG10}).

\subsection{Determination of the shock front and of the cobpoint}
\label{s2.3}

To identify the location of the shock front along the magnetic field line connecting the shock itself with the observer, we apply the same procedure as in \inlinecite{RG10}. We also use it to compute the  plasma variables and magnetic field values upstream and downstream of the shock front. 

Summarizing, the shock finder technique solves several steps. Starting from the position of the observer, and going back to the Sun, we search for the cobpoint location along the IMF line passing through the observer, by requiring a density or radial velocity value higher than a given threshold above the background solar wind. Then, we determine the direction of the shock normal, $\hat{\mathbf{n}}$, in order to characterize the shock. This requires to specify where the downstream region starts. We locate the downstream point as the first point after the cobpoint, going from the cobpoint location toward the Sun, where the velocity and the density start decreasing. The final step is to calculate the upstream-to-downstream ratio across the shock front, for the plasma variables of the simulation we are interested in\footnote{~Velocity, density and magnetic field ratios, or the angle between the magnetic field and the normal to the shock, $\theta_{Bn}$, for example.}. Here we concentrate in the normalized radial velocity jump, $VR = \frac{\upsilon_{r}(d)}{\upsilon_{r}(u)}-1$ (being $\upsilon_{r}(u)$ and $\upsilon_{r}(d)$ the radial velocity at the upstream and downstream point, respectively), because this is the variable used in the shock-and-particle models that allows us to derive a semi-empirical relation with the injection rate of shock-accelerated particles at the cobpoint, the \QVR relation \cite{Lario97,Lario98,Aran07a,Aran07b}.

\section{Evolution of plasma variables, magnetic field and \VR jump}
\label{s3}

\begin{figure*}[!t]
\centering
\includegraphics[width=1.0\textwidth]{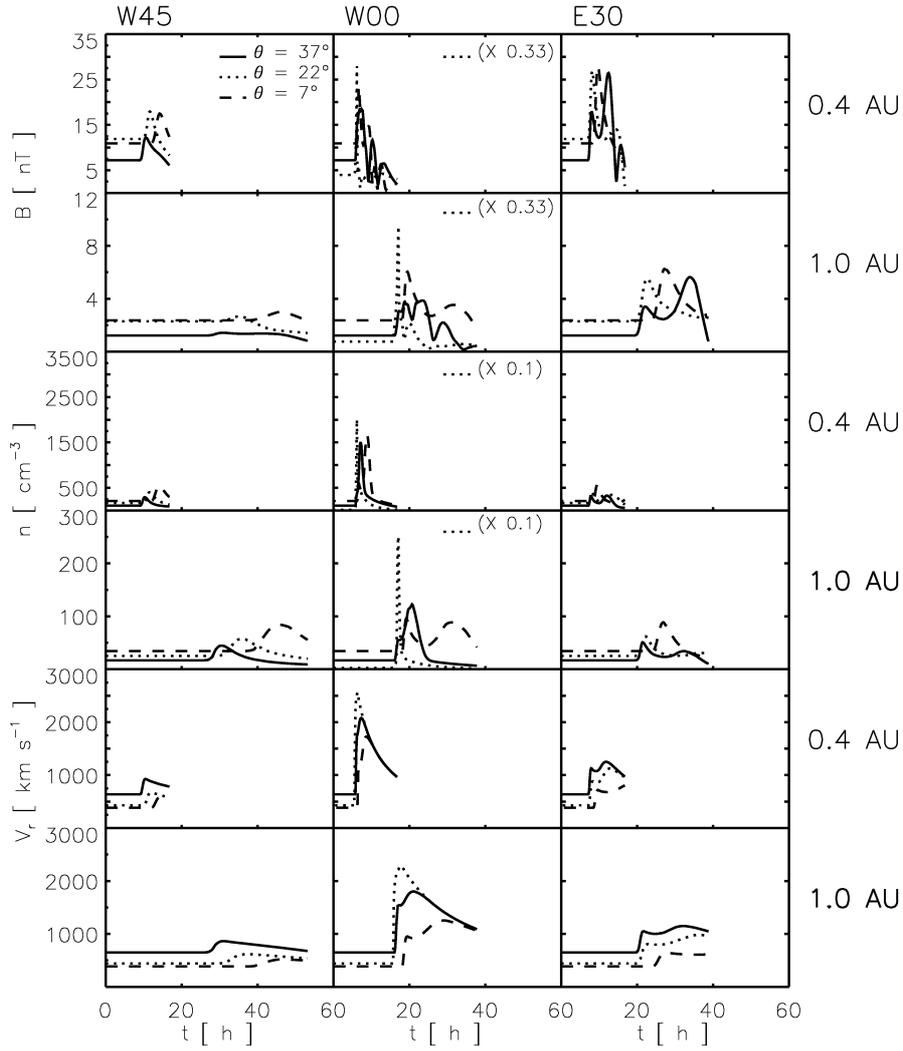}
\caption{From top to bottom: Evolution of $B$, $n$ and $\upsilon_{r}$ as seen by the nine 0.4\,AU-observers (upper panels) and 1.0\,AU-observers (lower panels). Each column shows the longitude of the observer and each different style line represents its latitude, as labelled. The dotted curves of the four top panels in the middle column had been reduced by the factor specified in each panel.}
\label{figplasma3d1}
\end{figure*}

From the IP shock propagation simulation we obtain the evolution of the solar wind plasma variables and magnetic field as measured for each of the nine observers located at 0.4\,AU and at 1.0\,AU, as shown in Figure~\ref{figplasma3d1}. Figure~\ref{figplasma3d1} displays, from top to bottom, the evolution of $B$, $n$ and $\upsilon_{r}$ for the observers situated at W45, W00 and E30 in longitude (left, middle and right panels, respectively). In each panel, the profiles corresponding to observers at different latitudes, N37, N22 and N07, are represented by the solid, dotted and dashed lines, respectively. The results for the 0.4\,AU-observers are thoroughly discussed in \opencite{RG10}, and they are included here for comparison.

Figure~\ref{figplasma3d1} shows that the increase in all plasma variables at the shock passage, the plasma jumps, are smaller at 1.0\,AU than at 0.4\,AU, because the shock becomes weaker as it expands into the IP medium. Nevertheless, in both cases the highest plasma jumps are those detected by the N22W00 observer. 

The times of the shock passage by each observer are listed in Table~\ref{tablecobcomparative}. The shock passage by the 0.4\,AU-observers occurs between 5.6 and 12.3\,hours after the launch of the perturbation, depending on the angular position of each observer. These times correspond to transit speeds (from the Sun to 0.4\,AU) between, approximately, 2900 and 1350\,km\,s$^{-1}$, respectively. In the case of the 1.0\,AU-observers, the shock arrives first to the W00 observers, being the shortest transit time for the N22W00 observer (15.9\,h, which implies a shock transit speed of around 2600\,km\,s$^{-1}$). Comparing with observed events at 1.0\,AU, this would be the second fastest recorded shock, staying between the 1972 August 4 event (14.6\,h of transit time) and the 1859 September 1 event (17.6\,h) \cite{Cliver04}. The shock arrives only 20\,minutes later to the N37W00 observer, while it takes 2\,hours more to pass by the N07W00 one. 

From Table~\ref{tablecobcomparative} we see that there is a noticeable delay of the shock passage among the 0.4\,AU-observers (up to 7\,hours) and among the 1.0\,AU-observers (up to 20\,hours), depending on their position. In general, for observers with the same longitude, the faster the background solar wind, the earlier the shock arrives to the observer, at both radial distances. These results mean that the shock arrives earlier to the $\theta$\,=\,37$^{\circ}$-observers in almost all the cases. The fastest shock, is however, seen by the N22W00 observer. For observers with the same latitude, the larger the separation between the shock nose and the observer, the later the shock passage.

\begin{figure*}[!h]
\centering
\subfloat{\includegraphics[width=0.65\textwidth]{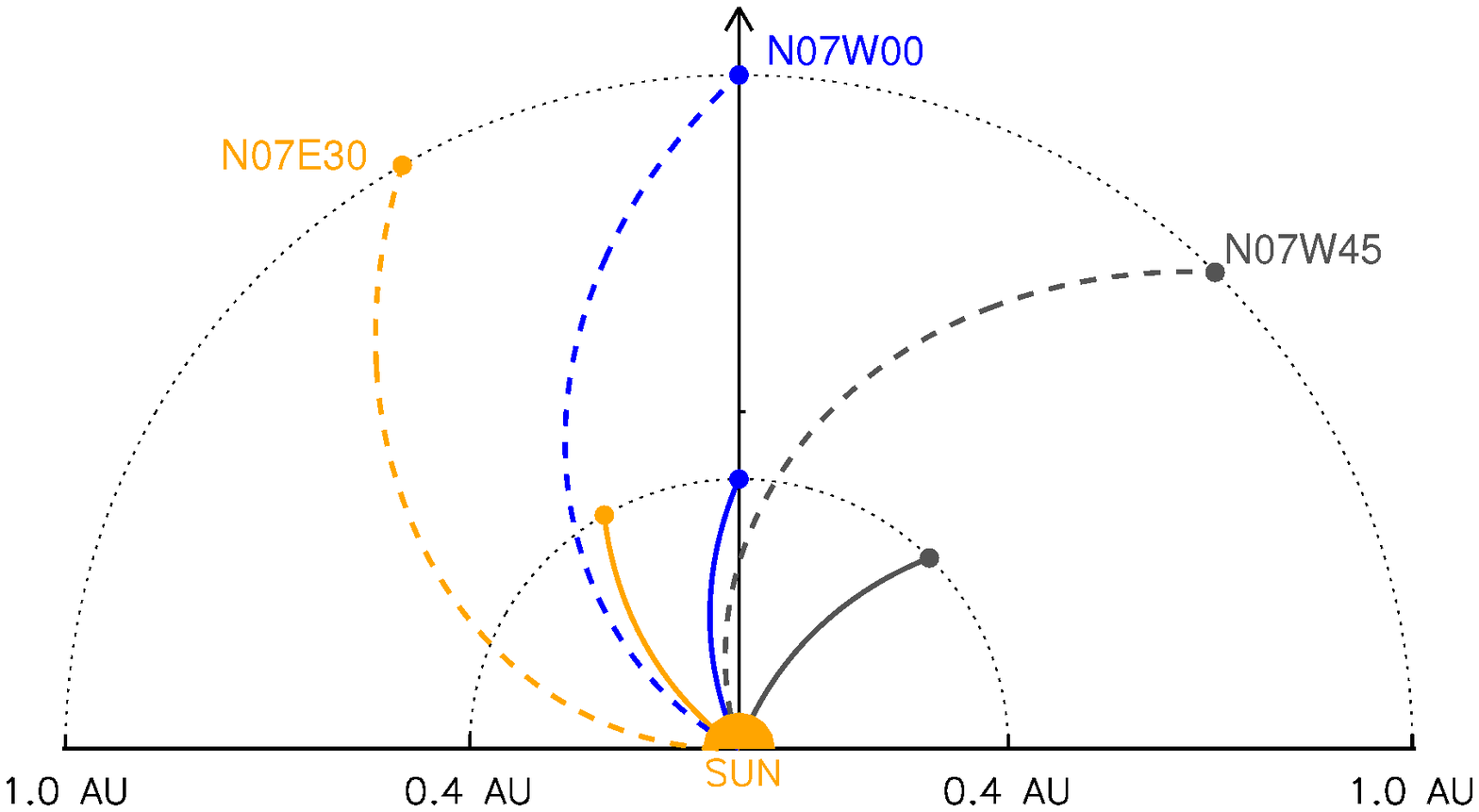}}\\
\subfloat{\includegraphics[width=0.65\textwidth]{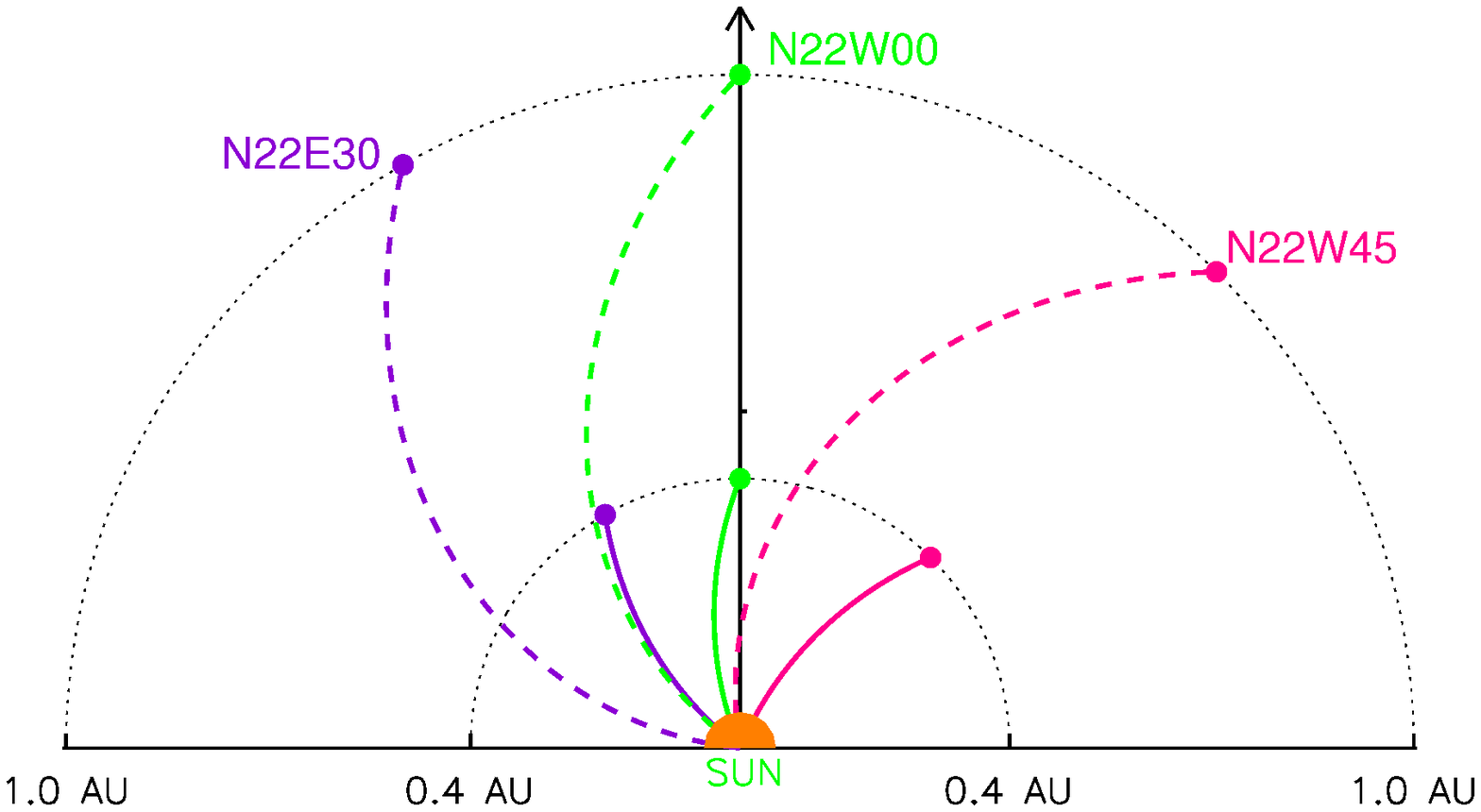}}\\
\subfloat{\includegraphics[width=0.65\textwidth]{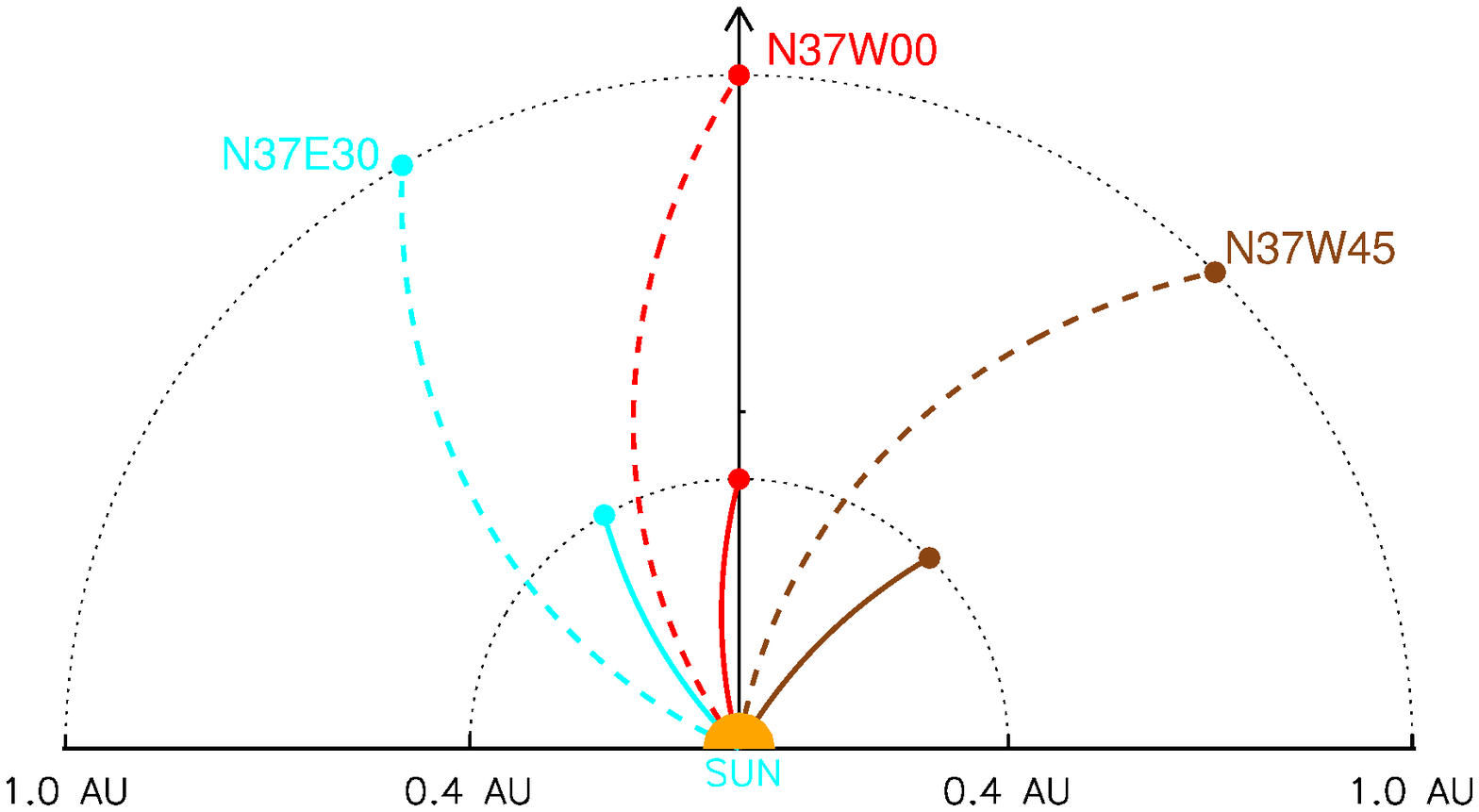}}
\caption{Comparison between the location of the first cobpoint for the 0.4\,AU- and the 1.0\,AU-observers. From top to bottom: $\theta$\,=\,7$^{\circ}$, $\theta$\,=\,22$^{\circ}$, and $\theta$\,=\,37$^{\circ}$. Each panel displays the position of the 0.4\,AU- and 1.0\,AU-observers and their corresponding IMF lines (solid and dashed lines, respectively). Observers are colour coded according to their position in space (see Section~\ref{s2.3}). The arrow indicates the main direction of the shock.}
\label{figfootpoint3d}
\end{figure*}

Figure~\ref{figfootpoint3d} compares the location of the first cobpoint for the 0.4\,AU- and the 1.0\,AU-observers, and their IMF lines. Columns (3) and (4) of Table~\ref{tablecobcomparative} lists the position of these first cobpoints. The differences in these cobpoints for observers located at the same longitude and latitude are due to the fact that the observers are placed at different radial distances, as well as to the variation of the solar wind speed with latitude. For example, the N22W00 1.0\,AU-observer lies on almost the same IMF line as the N22E30 0.4\,AU-observer, but this is not true for the other two latitudes. A consequence of these different magnetic connections is that, for observers with the same longitude and latitude but different radial distance, the evolution of \VR will not behave in the same way, because their respective cobpoints scan different regions of the shock front.

\begin{figure*}[!h]
\centering
\subfloat{\includegraphics[width=0.8\textwidth]{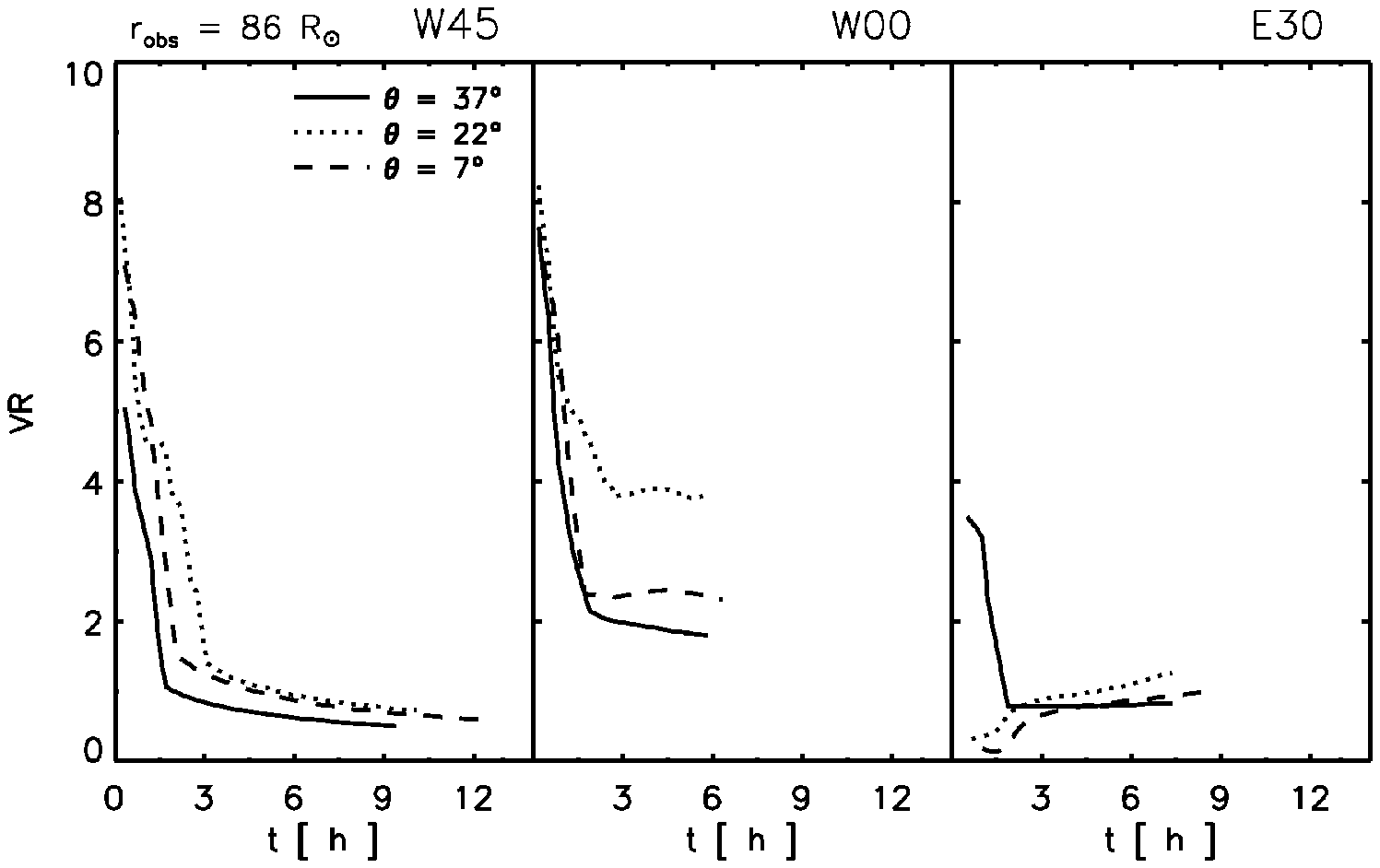}}\\
\subfloat{\includegraphics[width=0.795\textwidth]{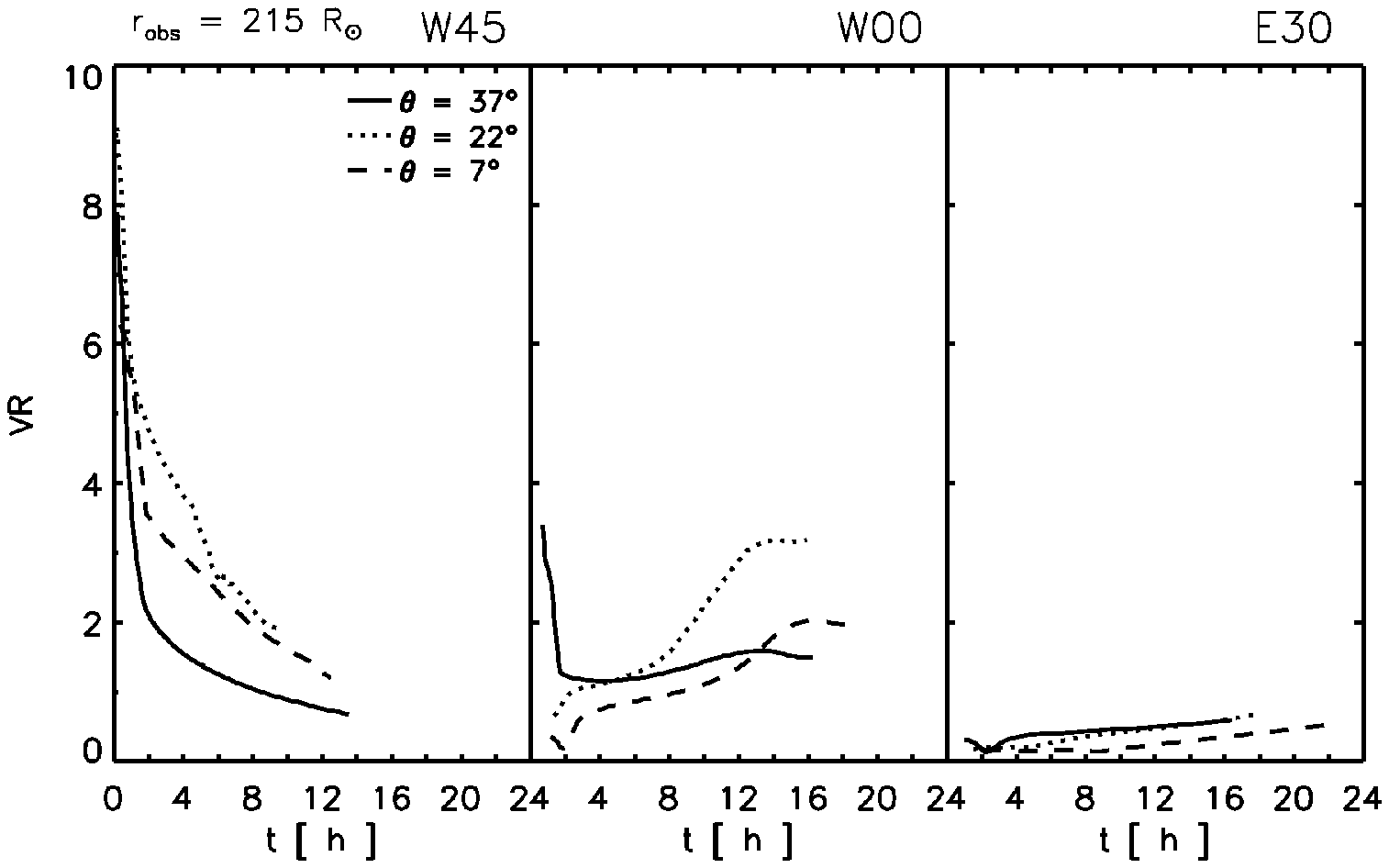}}
\caption{Evolution of \VR at the cobpoint for the nine 0.4\,AU-observers (top) and 1.0\,AU-observers (bottom). Each column shows the longitude of the observer and each style line represents its latitude, as labelled.} 
\label{figVR3d1}
\end{figure*}

Figure~\ref{figVR3d1} illustrates the evolution of \VR at the cobpoint for all the observers, while the shock is expanding up to 0.4\,AU (top) and 1.0\,AU (bottom). As in Figure~\ref{figplasma3d1}, the three panels show the evolution of \VR for the W45 (left), W00 (middle) and E30 (right) observers. The solid, dotted and dashed lines correspond to the N37, N22 and N07 latitudes, respectively. For the 0.4\,AU-observers, the highest \VR value is reached by the N22W00 observer, whereas for the 1.0\,AU-observers by the N22W45 observer. These observers are the ones connected closest to the shock nose at the beginning of the event, at N22E22 and at N22E07, respectively. Since this latter connection is closer to the shock nose, the N22W45 1.0\,AU-observer measures the greatest value of \VR.

\begin{center}
\begin{table}[!h]
\caption{Comparison between the shock passage time ($t_{s}$) and the location of the first cobpoint for the 0.4\,AU-observers (left) and the 1.0\,AU-observers (right).}
\label{tablecobcomparative}
\begin{tabular}{llrrrr}
\hline
  & &  \multicolumn{2}{c}{$t_{s}$\,[h]} &  \multicolumn{2}{c}{First cobpoint}\\
  & & 0.4\,AU & 1.0\,AU & 0.4\,AU & 1.0\,AU\\
\hline
\multirow{3}{*}{W45} & $\theta = 7^{\circ}$ & 12.3 & \textit{37.0}* & N07W22 & N07E13 \\
 & $\theta = 22^{\circ}$ & 10.1 & \textit{29.0}* & N22W23 & N22E07 \\
 & $\theta = 37^{\circ}$ & 9.4 & \textit{26.3}* & N37W30 & N37W07 \\
\hline
\multirow{3}{*}{W00} & $\theta = 7^{\circ}$ & 6.3 & 18.1 & N07E23 & N07E59 \\
 & $\theta = 22^{\circ}$ & 5.6 & 15.9 & N22E22 & N22E53 \\
 & $\theta = 37^{\circ}$ & 5.8 & 16.2 & N37E16 & N37E38 \\ 
\hline
\multirow{3}{*}{E30} & $\theta = 7^{\circ}$ & 8.5 & \textit{24.2}* & N07E53 & N07E88 \\
 & $\theta = 22^{\circ}$ & 7.3 & \textit{19.9} & N22E52 & N22E84 \\
 & $\theta = 37^{\circ}$ & 7.3 & \textit{19.8} & N37E46 & N37E68 \\
\hline
\end{tabular}
~\\
\begin{footnotesize}
(*~For the 1.0\,AU-observers, the time given in italic refers to \\
the perturbation passage, since the discontinuity is too\\
weak to be considered as a shock.)
\end{footnotesize}
\end{table} 
\end{center}

In the magnetically well-connected cases, there is a rapid decrease of the values of \VR within, approximately, the first two hours of the simulation. This is the case, for example, of the N22W00 observer at 0.4\,AU, which is well-connected from the beginning of the event (N22E22), and so the high values of \VR; but in spite of the movement of its cobpoint toward the nose of the shock, \VR decreases down to a constant value. These high-values of \VR obtained close to the Sun may be due to a combination of several factors, as the fact that the solar wind speed is still increasing\footnote{~For example, for t = 0.5\,h, the cobpoint is located at 10.3\,\Rsol, the upstream solar wind speed is 300.4\,km\,s$^{-1}$, and $ \upsilon_{r}(d)$ = 2225.6\,km\,s$^{-1}$, yielding \VR = 6.38.}, that at these distances ($r$\,$\leq$\,15\,\Rsol) the shock has not yet detached from the actual compressed material forming the `ejecta' of the CME itself (as found in other simulation schemes, \opencite{Manchester08}; \opencite{Luhmann10}; \opencite{RG10}), as well as due to the deceleration of the shock. In contrast, the same observer at 1.0\,AU magnetically connects to 31$^{\circ}$ farther away from the nose of the shock (N22E53), and, hence, its cobpoint scans the left wing of the front, moving toward more central positions. This is the reason of the low values of \VR at the beginning of the event, which increase monotonically with time. These values are smaller than for the 0.4\,AU-observer because the 1.0\,AU-observer is connected to the weak wing of the shock when it is still strong (\textit{i.e.}, close to the Sun), and because its cobpoint only reaches the central region when the shock has travelled a longer distance. Except for one 0.4\,AU-observer, the E30 observers connect to the eastern wing of the shock, which weakens as it expands; consequently, their \VR evolution is practically constant and smaller than in the other cases discussed in \cite{RG10}.

\section{Flux profiles and radial variation}
\label{s4}

In order to visualize and quantify the influence of the latitude of the observer on SEP events, we produce synthetic flux profiles as measured by the virtual observers described in Section~\ref{s2}, by using the \QVR relation of the SOLPENCO code \cite{Aran06}. As we assume a functional dependence $\log\,Q$\,$\propto$\,\VR, the evolution of \VR for different observers directly translates into an evolution of the injection rate of shock-accelerated particles, and, therefore, in a variety of SEP flux profiles. The values adopted for the description of the IP particle transport and other features of the model can be found in \inlinecite{Aran06}.

\subsection{Synthetic flux profiles}
\label{s4.1}

\begin{figure}[!h]
\centering
\subfloat{\includegraphics[width=0.775\textwidth]{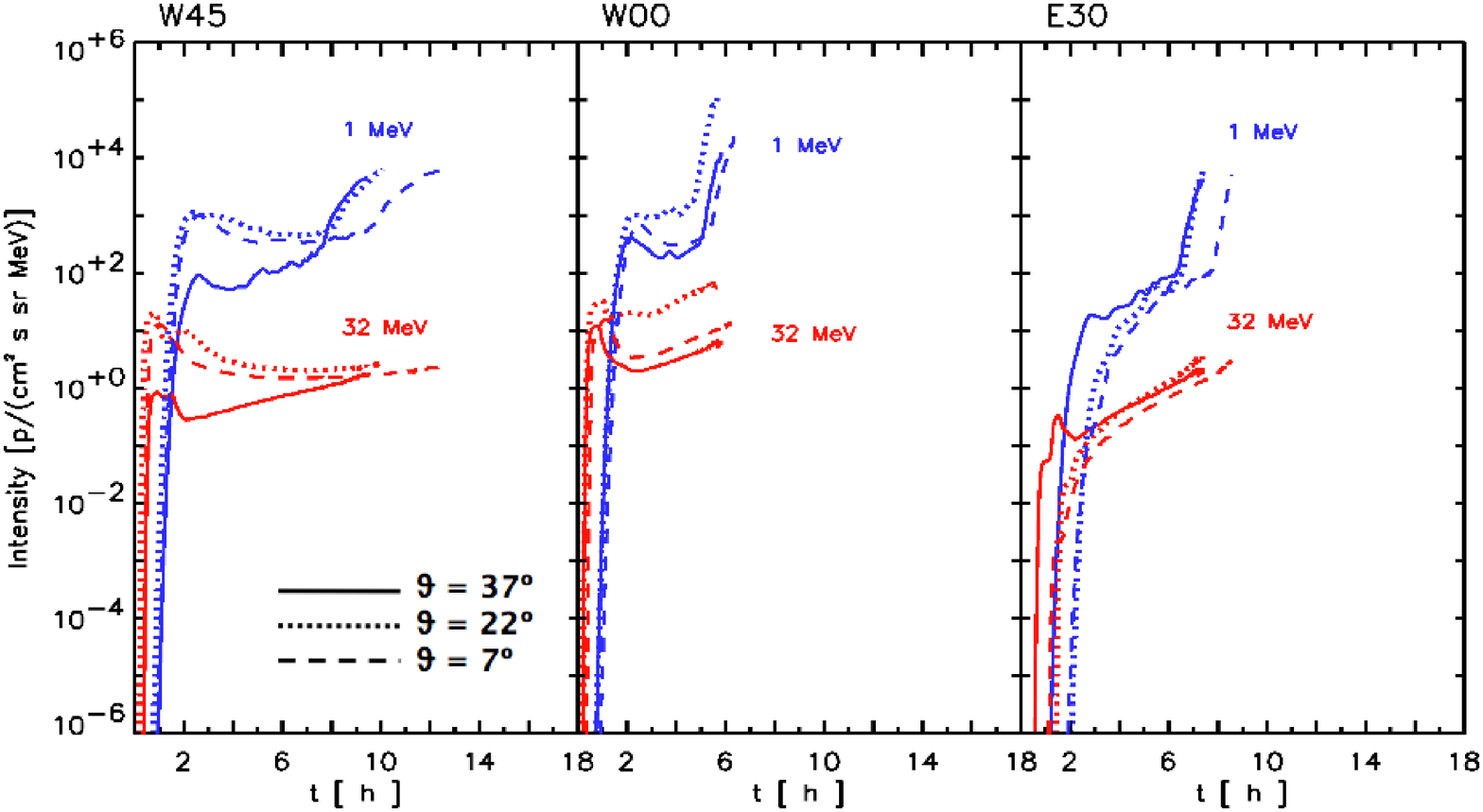}}\\
\subfloat{\includegraphics[width=0.775\textwidth]{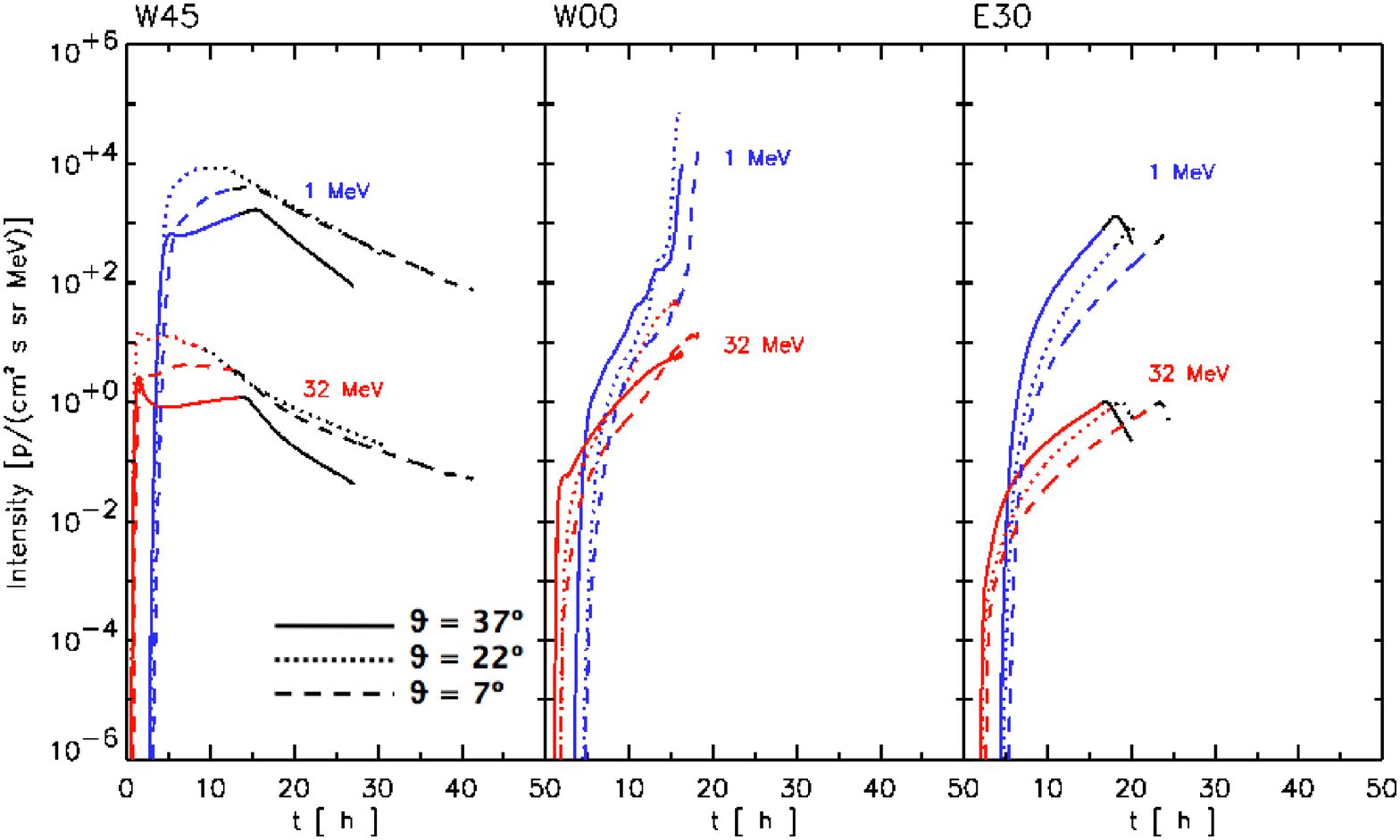}}
\caption{Simulated flux profiles derived from the particle transport model for 1\,MeV (blue lines) and 32\,MeV (red lines) protons, for the nine 0.4\,AU-observers (top) and 1.0\,AU-observers (bottom). Each column shows the observer's longitude and each style line represents its latitude, as labelled.}
\label{figfluxes3d1}
\end{figure}

Figure~\ref{figfluxes3d1} shows two sets of synthetic flux profiles derived for the nine 0.4\,AU-observers (top) and 1.0\,AU-observers (bottom). Blue lines are the derived flux profiles for 1\,MeV protons, with a mean free path that depends on the rigidity of the particle as $R^{+0.5}$, assuming a region of enhanced pitch-angle scattering ahead of the shock, a foreshock, in order to simulate the energetic storm particle (ESP) component of SEP events often seen at low-energies ($<$\,5\,MeV) at the crossing of the shock by the observer; see more details in \inlinecite{Aran06} and references therein. Red lines are the flux profiles at 32\,MeV, with the same expression for the mean free path and without considering any foreshock. In the case of the W45 and E30 1.0\,AU-observers, the injection rate has been exponentially decreased five orders of magnitude (these portions of the flux profiles are drawn in black in Figure~\ref{figfluxes3d1}), when the shock becomes too weak or it is tagged as ``not a shock'' by the shock finder algorithm. Therefore, since the shock front arriving to these western and eastern observers is weak, we have only simulated a foreshock for the central meridian (W00) observers at low energy.

The derived synthetic proton flux profiles are a consequence of the continuous injection of shock-accelerated particles, controlled by \VR, plus the conditions of the particle transport along the IMF lines. The maximum intensity is measured by the N22W00 observer at the shock passage in the two radial distances considered, both at 1\,MeV and at 32\,MeV. The shape of the proton flux profiles for the corresponding observers at different radial distances might differ considerably: for example, at 1\,MeV, for the W00 0.4\,AU-observers, there is a sudden increase of the intensity at the prompt phase, followed by a further significant increase at the shock passage (due to the foreshock region). For the W00 1.0\,AU-observers, however, the flux profiles evolve more smoothly up to the shock arrival. As expected from the values of \VR, the prompt phase in the case of the W00 and E30 observers at 0.4\,AU is more intense than in the case of the corresponding 1.0\,AU-observers; the reason is, once more, that the magnetic connections of the 0.4\,AU-observers are established closer to the shock nose and earlier than for the 1.0\,AU-observers. For the well-connected events, \textit{i.e.}, W45 cases and W00 at 0.4\,AU, the peak intensity at high-energy is attained at the prompt phase (except for N22W00, where the nose of the shock crosses the observer), in agreement with observations. This is an improvement with respect to the intensity-time profiles provided by SOLPENCO at 1.0\,AU, for which peak intensities were often attained at the shock passage even at high energies rendering, therefore, the time occurrence prediction of the peak intensities of SOLPENCO valid only for E\,$<$\,5\,MeV \cite{Aran08}. We find these peak intensities at the prompt phase because we can trace the magnetic connection of the observer with the shock front when it is moving close to the Sun ($\sim$\,4\,\Rsol), allowing us to simulate the early injection of particles by the shock. Similar results are also found by \cite{Aran11} following a 2D MHD shock simulation approach.  

The presence of a foreshock that confines low-energy particles might have a significant effect on the shape of the flux profiles. With an active foreshock at 1\,MeV, for example, the maximum intensity at the prompt phase is exceeded by the intensity value at the shock passage. But this depends on the energy of the particles, the radial distance of the observer and on its angular (longitude and latitude) position with respect to the leading edge of the shock. For the 0.4\,AU-observers, the intensity peaks at the shock passage, and it is higher as the observer's position is closer to the main shock direction. 

In conclusion, the variation of the effects of the foreshock region with the radial distance depends mainly on the characteristics of the modelled shock, and on the way we simulate this foreshock. Nevertheless, it is risky to extend the conclusions about the influence of the foreshock only from the analysis of these simulations. In other scenarios (a wider shock, for example), the 1\,MeV-intensity profile (and even the 32\,MeV one) for the E30 1\,AU-observers could display the peak intensity at the shock passage; this is the case of the 1991 March 24 ESP event \cite{Lario08}, for example. It is worth to remember that a consistent IP shock-foreshock model, contrasted with observations, does not yet exist (at different radial and angular distances). However, efforts to include such contribution within simulations are being made \cite{Vainio07b,Vainio08,Battarbee11}.

\subsection{Radial variation of peak intensities}
\label{s4.2}

\begin{figure}[!t]
\centering
{\includegraphics[width=0.995\textwidth]{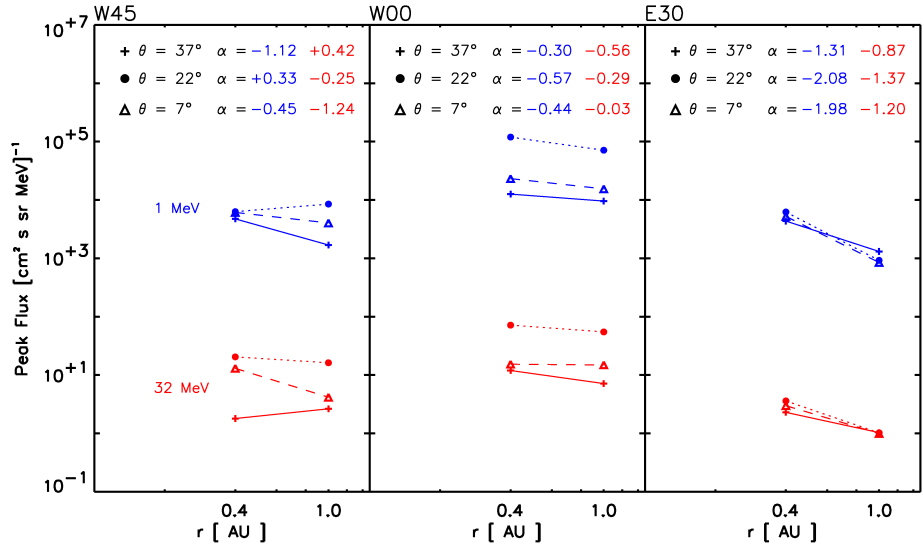}}\\
\caption{Radial variations of 1\,MeV (blue) and 32\,MeV (red) proton peak intensities for the 0.4\,AU- and the 1.0\,AU-observers. The values of $\alpha$ are given in the inset for each latitude and energy.}
\label{figdepr3d3}
\end{figure}

Comparing the peak intensity values obtained at 0.4\,AU and at 1.0\,AU, we can derive a radial index, $\alpha$, assuming that the peak intensity, $P$, varies with the radial distance as $P$\,$\propto$\,$r^{\alpha}$ (following the same approach as \opencite{Aran07a}; \opencite{Lario07}; \opencite{Verkhoglyadova12}).  Figure~\ref{figdepr3d3} presents the peak intensity values at both radial distances, as well as the derived indices.

Allowing for small variations, the general tendency of the radial dependence is similar for both energies for a given longitude (\textit{i.e.}, the tendency is the same in latitude), but it largely changes for observers at different longitudes. In general, the peak intensity decreases with radial distance, except for the observers N22W45, at low energy, and N37W45, at high energy. The reason is that these W45 observers at 1.0\,AU have the best connection with the shock at the beginning of the event (see Table~\ref{tablecobcomparative} and Figure~\ref{figfootpoint3d}), and so the peak intensity is achieved at the prompt phase of the event (remember that the shock does not reach 1.0\,AU); while the 0.4\,AU-observers establish their magnetic connection in a region where \VR is smaller than for the 1.0\,AU-observers (see Figure~\ref{figVR3d1}) resulting in intensity-time profiles peaking at the shock passage, where the shock is still efficient at accelerating particles. If the 0.4\,AU-observers were placed in the same magnetic flux tube as the N22W45 and N37W45 1.0\,AU-observers, \textit{i.e.}, sharing the same magnetic connection with the shock front, peak intensities would decrease with radial distances as $r^{-2.61}$ and  $r^{-2.00}$ for 1\,MeV and 32\,MeV, respectively, for the N37W45 case, and $r^{-2.77}$ and  $r^{-2.02}$ for low and high energy, respectively, for the N22W45 case. Therefore, radial variations of peak intensities strongly depend on the way the observers are connected with the shock front and on how efficient is the shock in the regions scanned by their cobpoints, specially for a narrow and rapid shock like the one simulated in this work. 

For the W00 and E30 observers, the radial indices both at low and high energy follow the same trend. For central meridian observers, the peak intensities softly decrease with radial distance because the shock nose is still efficient at accelerating particles when it crosses 1.0\,AU. In the case of the eastern observers, peak intensities decrease faster than for W00 observers owing to the smaller values of \VR scanned by the cobpoints of the 1.0\,AU-observers with respect to those of the 0.4\,AU-observers (see Figure~\ref{figVR3d1} and discussion in Section~\ref{s3}). Note that the decrease of the peak intensity with radial distance is softer for the N37E30 case because the observer at 1.0\,AU has a better connection than those at N22 and N07.

The wealth of observational data that allows us to compare with the radial dependences that we find come from simultaneous observations of SEP events from the \textit{Helios-1} and \textit{Helios-2} spacecraft (from 0.29 to 0.98\,AU) and the near-Earth observatories \textit{IMP-8} and \textit{ISEE-3}. A number of multi-spacecraft studies have been performed using these data sets (see a detailed summary in \opencite{Lario13}, and references therein). We focus here in those studies providing radial dependences for the maximum intensity of protons at the energies between 1\,MeV and 32\,MeV. In their Figure 4, \inlinecite{McGuire83} show the radial dependences of six SEP events for 11\,--\,60\,MeV protons. Among these events, whose magnetic connections were within 35$^{\circ}$ from the parent event site, two had the \textit{Helios} spacecraft located at $r$\,$<$\,0.6\,AU. The radial dependences derived for these events are $\alpha$\,$\sim$\,0.09 and $\alpha$\,$\sim$\,-1.8. These events are comparable with our simulations for 32~MeV protons for the W45 and N37W00 cases (see their magnetic connections in Table~\ref{tablecobcomparative}) for which we obtain -1.24\,$\leq$\,$\alpha$\,$\leq$\,+0.42 (see Figure~\ref{figdepr3d3}) and -2.02\,$\leq$\,$\alpha$\,$\leq$+0.79 when considering 0.4\,AU-observers sharing the same IMF line as the 1.0\,AU-observers. \inlinecite{Lario06} analysed 72 SEP events seen simultaneously by at least two spacecraft (among the two Helios and IMP-8) at two different energies, 4\,-\,13\,MeV and 27\,--\,37\,MeV protons. These authors found that for these events the peak intensity has a radial index $\alpha$\,=\,-2.09\,$\pm$\,0.23 for 27\,--\,37\,MeV which is steeper than the values we find for 32~MeV protons. Nevertheless, \inlinecite{Lario06} show that the variation of the peak intensity with radial distance largely changes from event to event (\textit{e.g.} $r^{-2.32}$ for the event on doy 118 in 1978, and $r^{-0.23}$ for the event on doy 169 in 1981).

The \QVR relation used in the SOLPENCO tool \cite{Aran06} is applied to obtain synthetic proton intensity-time profiles in the energy range of 0.125\,--\,64~MeV for 28 virtual observers placed on the ecliptic plane and located at 0.4 and 1.0 AU with longitudes extending from W90 to E65 (including W45, W00 and E30). The synthetic SEP events in SOLPENCO are produced by simulating 8 interplanetary shocks (from 18\,\Rsolar to 1.1\,AU) with different transit speeds to 1.0\,AU (from 25 to 62~hours for the W00 observer). We refer the reader to \inlinecite{Aran07a} for more details. Since the parameters defining the proton injection rate by the shock and the particle transport are the same as in SOLPENCO and the virtual observers are also placed at the same radial and longitudinal positions, any difference in the obtained radial indices is mainly due to the different shocks simulated, and to the different solar wind speeds used for different latitudes (affecting the convection and adiabatic deceleration of the particles, specially at low-energies). For W45, W00 and E30 observers, SOLPENCO yields -1.4\,$<$\,$\alpha$\,$<$\,0.6 at the two energies considered. We obtain similar values for the W45 and W00 observers, but not for the eastern cases for which we obtain -2.08\,$<$\,$\alpha$\,$<$\,-0.87. The shocks simulated in SOLPENCO are wider than the narrow shock simulated in this work, resulting in a better magnetic connection for the eastern 1.0\,AU-observers with the shock front than in our case. This favours a softer decrease of the peak intensity with the radial distance. 

\inlinecite{Verkhoglyadova12} use the PATH model to simulate the shock propagation (from 0.1 to 2.0\,AU) and the acceleration and transport processes of protons in the energy range of 0.3\,-\,5\,MeV/nuc, and of iron ions. These authors compute the particle intensity-time profiles as seen by four virtual observers located at 0.5, 1.0, 1.5 and 2.0\,AU. By assuming different conditions for the shock-acceleration processes (\textit{i.e.}, different shock obliquities, with $\theta_{Bn}$\,=\,15$^\circ$, 45$^\circ$ and 75$^\circ$) and different compositions for flare and supra-thermal solar wind seed particle populations (\textit{i.e.}, flare-only, mixed, and shock-only cases), they obtain radial variations for the 1\,MeV protons, ranging approximately from $\alpha$\,=\,-2.9 to $\alpha$\,=\,-1.7 (see their Figure~4a) with increasing values of $\alpha$ up to 5\,MeV. The PATH model considers the particle acceleration processes; however, at present, PATH simulations are restricted to 1D modelling and, therefore, it cannot account for the longitudinal (and latitudinal) variations of peak intensities. This limitation makes it difficult to compare the radial dependences derived from the PATH model with those derived from our shock-and-particle model, even if we track both the radial and the longitudinal variation of the shock-particle injection conditions along the region of the shock front where the observer is connected to (and also, by considering observers at different latitudes, the variations in latitude). In order to establish some comparison between the two models, we can consider those cases where we locate the 0.4\,AU- and the 1.0\,AU-observers at the same magnetic flux tube, as they do. For 1\,MeV protons, we find -2.77\,$\leq$\,$\alpha$\,$\leq$\,-2.20 for the well-connected cases (W45), and -1.63\,$\leq$\,$\alpha$\,$\leq$\,-1.07 for the eastern cases, poorly connected. The radial dependence of the peak intensity is similar to that of \inlinecite{Verkhoglyadova12} in the case of the W45 observers. This is an expected result since these peak intensities are reached very early in the event, at the prompt phase of the flux profiles at 1.0\,AU, when the cobpoint of the observers has not yet moved from the central region of the shock front.

In the case of the central meridian observers (W00), we obtain that peak intensities increase with the radial distance, $\alpha$\,=\,+0.42 for N37W00 and $\alpha$\,=\,1.37 for N07W00. The increase is specially large in the case of the N22W00 observers, $\alpha$\,=\,4.74, because the nose of the fast (and narrow) shock that we have modelled crosses the 1.0\,AU-observer, whereas the 0.4\,AU-observer is only reached by the western wing of the shock, at $\sim$\,30$^{\circ}$ away from the nose, and thus the lower peak intensity attained at 0.4\,AU (compare the events N22E30 at 0.4\,AU and N22W00 at 1.0\,AU in Figure~\ref{figfluxes3d1}). It is worth noting that our model does not provide a prediction for the downstream region of the events, where presumably in an event like the one seen by the N22W00 observer, the peak intensity at 0.4\,AU would have been attained in the downstream, so the actual situation would be a softer radial dependence than what we find here.  
 
It is concluded from our simulations and previous modelling efforts assuming a continuous contribution of shock-accelerated particles \cite{Aran05,Ruzmaikin05,Vainio07,Aran11,Verkhoglyadova12}, that radial dependences of peak intensities (and fluences) in SEP events depend on several factors: the particle energy, the efficiency of the shock at accelerating and injecting particles, the available seed particle population, the obliquity of the shock and the changing conditions scanned at the shock front by the cobpoint. This variety of factors makes it difficult (if possible) to derive a universal radial dependence valid for all types of gradual SEP events. Besides, the observational study by \inlinecite{Lario06} concludes that the longitudinal angular separation between the observer and the solar parent activity is the fundamental parameter that controls the radial variation of the peak intensity. In addition to all these factors, we add the latitudinal position of the observer as another parameter to be considered when deriving the radial variation of peak intensities in SEP events. 

\section{Conclusions} 
\label{s5}

The evolution of the magnetic field intensity, number density and radial velocity for several observers located at 0.4\,AU and at 1.0\,AU, and at different heliolongitudes and heliolatitudes has been analysed, discussing the relevance of the latitude of the observer. We have found that the highest plasma jumps at the shock passage are detected by the observer located in the direction of the shock nose, \textit{i.e.}, the N22W00 observer in our simulation setup, at both distances, and that there is a considerable reduction in the plasma values as the observer is placed far apart from the shock nose, both in longitude and in latitude. The general behaviour of the plasma and magnetic field jumps are qualitatively the same at 0.4\,AU and at 1.0\,AU, being the jumps smaller at 1.0\,AU.

The evolution of the normalized radial velocity jump, \VR, shows that: (1) MHD simulations of shocks have to include the evolution of the plasma variables and magnetic field close to the Sun, since the strength of the shock may rapidly decrease with radial distance and, hence, the efficiency of the shock as particle accelerator (which is especially important at high energies); and (2) \VR also varies with the latitude, not only with the longitude. The highest values of \VR correspond to those observers with cobpoints nearer to the shock nose, their location depending on the observer0s position and latitude. Therefore, combined models of shock propagation and particle acceleration, injection and transport, should take into account the influence of the variations of the plasma variables at the shock front not only with radial distance and longitude, but also with latitude.

We have derived the corresponding synthetic flux profiles within the frame of our shock-and-particle model \cite{RG11}, and we have presented examples illustrating the relevance of the observer's latitude. It has been shown that, for the two energies studied, the N22W00 observer gathers the maximum intensity. Within the same shock simulation, the flux profiles can differ up to one order of magnitude for observers with the same longitude but different latitudes, for both radial positions. These differences imply a large variation in the efficiency of the shock at injecting accelerated particles, being larger for the 0.4\,AU-observers than for the 1.0\,AU ones.

Because the cobpoint is traced back closer to the Sun than in former models, the injection of the high energy particles near the Sun can be consistently simulated allowing us to reproduce the prompt phase of gradual well connected SEP events. We find that the intensities attained at the prompt phase depend on the observer's magnetic connection to the shock front (which in turns depends on the latitude), in such a way that the better the connection between the observer and the shock at the beginning of the event, the more intense the prompt phase. In general, the prompt phase is more intense for the 0.4\,AU-observers than for the 1.0\,AU ones. 

Regarding the peak intensities, they largely change up to one order of magnitude for observers at any distance or longitude, depending on their latitude. We have found that the N22W00 observer is the one who attains the maximum intensity in all scenarios. It has been noted, however, that the presence of a foreshock can modify the shape of the flux profiles and, so, the peak intensity values and their time occurrence. At high energy, the peak usually appears at the prompt phase; but at low energy, it could be exceeded by the value reached at the shock passage. For those peaks that occur at the prompt phase of the event, the better the magnetic connection between the observer and the front shock, the higher the peak; while for the cases when the peak intensity occurs at the shock passage, the closer the observer to the shock nose the larger the peak.

From the study of the radial dependence of the peak intensity, we obtain that in general it decreases with radial distance. The only exceptions are the N22W45 observers at low energy and the N37W45 observer at high energy, due to the best connection with the shock at the beginning of the event of the W45 1.0\,AU-observers. The W00 and E30 observers present a similar trend with the observers' latitude. For W00 observers, peak intensities decrease softly due to the contribution of the shock, and for eastern cases the radial dependences are steeper because of the poor connection. All these results are a consequence of the way each observer is connected to the front of this narrow shock.

The main conclusion of this study is the relevance of the latitude of the observer with respect to the leading direction of the shock. This is a factor scarcely commented and quantitatively rarely addressed in numerical simulations of SEP events. At present, practically all efforts have been focused on the longitude of the observer, principally because: (1) the main body of observations comes from spacecraft located near 1.0\,AU close to the ecliptic plane; and (2) 3D modelling of SEP events (including the simulation of the CME-driven shock) is a complex and computer demanding task hardly affordable, even nowadays.

We maintain that the inclusion of the latitude is important for space weather purposes and, therefore, it deserves further attention. Needless to say that many more simulations (as other radial distances, longitudes and latitudes, shock velocities and shapes, etc.) are required to draw definitive conclusions. 

\begin{acks}
These results were obtained at Universitat de Barcelona in the framework of the projects AYA2007-60724 and AYA2010-17286 of the Spanish Ministerio de Ciencia e Innovaci\'{o}n. A.~A. was supported by an ESA internal research fellowship. R.~R-G. also acknowledges economical support from EU FP7 SEpserver project (n. 262773). C.~J. and S.~P. acknowledge the support of the GOA/2009-009 (K.U.~Leuven), G.0729.11 (FWO-Vlaanderen) and C~90347 (ESA Prodex 9), and partial funding from the EU FP7 projects SOLSPANET (n. 269299), SPACECAST (n. 262468), eHeroes (n. 284461) and SWIFF (n.263340). The authors acknowledge the use of the computing facilities of the supercomputer cluster of CESCA (Centre de Serveis Cient\'{\i}fics i Acad\`{e}mics de Catalunya, Barcelona) and the high performance VIC cluster at the Katholieke Universiteit (Leuven). We are also grateful to the faculty of ESA (European Space Agency) for the financial support to visit ESA/ESTEC.
\end{acks}


\end{document}